\documentclass[12pt]{article}
\usepackage[utf8]{inputenc}
\usepackage[english]{babel} 
\usepackage[dvips]{graphicx}
\usepackage[round,authoryear]{natbib}
\usepackage{float}
\usepackage{perpage}
\usepackage{amsmath}
\usepackage{amsfonts}
\usepackage{multirow}
\usepackage{rotating}
\usepackage{hyperref}
\MakeSorted{figure}
\MakeSorted{table}
\graphicspath{{images/}}
\newcommand{\angstrom}{\textup{\AA}}
\newcommand{\footremember}[2]{%
    \footnote{#2}
    \newcounter{#1}
    \setcounter{#1}{\value{footnote}}%
}
\title{Accuracy of the component mass estimation methods in LMXB.}
\author{Antokhina $^1$\footremember{alley}{elant@sai.msu.ru}%
 E. A., Petrov$^1$ V. S., Cherepashchuk$^1$ A. M.}
\date{%
    $^1$Sternberg Astronomical Institute, Moscow M.V. Lomonosov State University\\[2ex]%
    \today}
\begin{document}
\maketitle
\begin{abstract}
There is a mismatch between modelled and observed distributions of optical stars masses in BH LMXB.
Companion masses in BH LMXB are found in the mass range 0.1 - 1.6 $M_{\odot}$ with the peak at 0.6 $M_{\odot}$. The standard evolutionary scenarios require the donor mass distribution peaks to be at least $\sim 1\; M_{\odot}$ to eject a massive envelope of the black hole progenitor. Imperfect of the methods of optical stars masses determination may cause this difference.

Two common used approximations of real Roche lobe filling star as distortion sources have been tested. On the one hand, there is the a approximation of real Roche lobe filling optical stars as sphere. We tested rotational broadening of absorption lines based on an exact calculation of CaI $\lambda 6439.075 \; \angstrom $ absorption profiles in the spectra of Roche lobe filling optical stars. There is overestimation of  projected equatorial rotational velocity $V_{rot} \sin i$ and, accordingly, underestimation of mass ratio $q=M_x/M_v$ in the spherical star model. On the other hand, the approximation of a real Roche lobe filling star as disk with uniform local profile and linear limb darkening law is more rough. In this case overestimation of  projected equatorial rotational velocity $V_{rot} \sin i$ is $ \sim  20\%$. Such overestimation of $V_{rot} \sin i$ can result in significant underestimation of mass ratio $q=M_x/M_v$ at hight value of $q=M_x/M_v$. Refined value of $q$ does not affect the mass of a black hole, but  the mass of an optical star has shrunk 1.5 times. Hence, the masses of optical stars in BH LMXB must be corrected downward, despite the contradictions to the standard evolutionary scenarios.
\end{abstract}

\section{Introduction}

\begin{figure}[H]
\center{\includegraphics[width=100mm,scale=0.6]{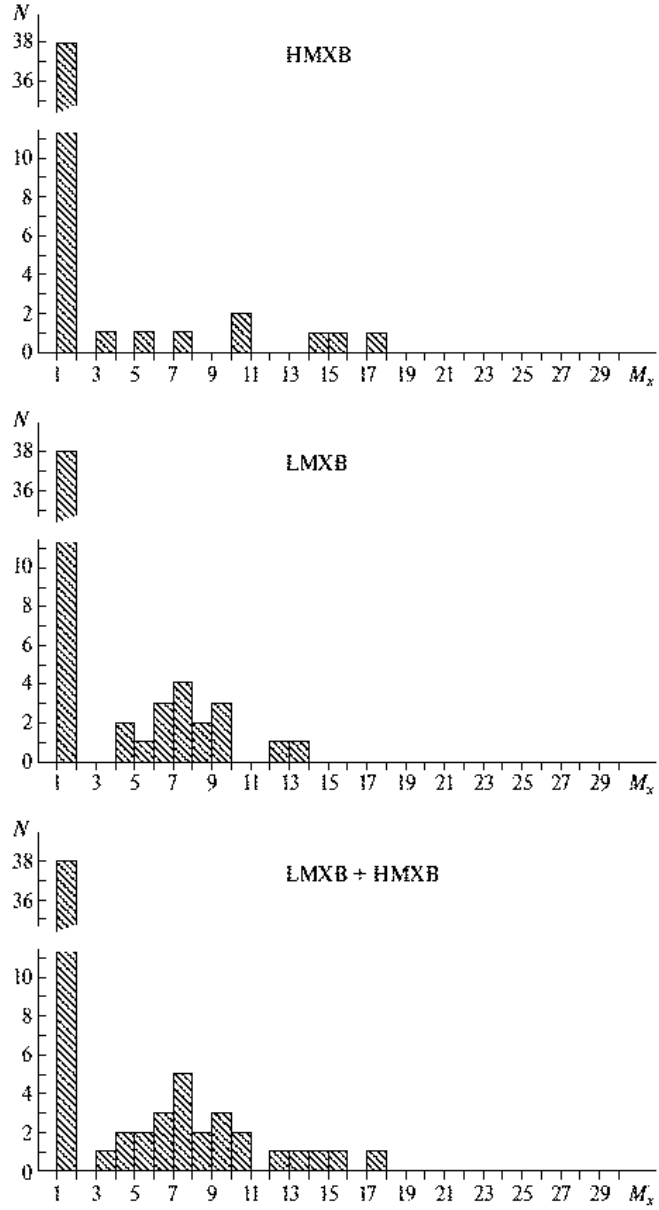}}\\
\caption{Mass distributions for neutron stars (high peak to the left) and black holes in high-mass and low-mass X-ray binaries.}
\label{ris:Distrb}
\end{figure}

There are 26 currently known X-ray binary systems with black holes : 17 with low mass  optical companions (LMXB) with $M_v = 0.3 \div 2 M_{\odot}$ and 9 with high mass optical companions (HMXB) with $M_v = 5 M_{\odot} - 70 M_{\odot}$  \citep{AnMihKniga}. Evolution scenarios for this two groups are completely different. Evolution of BH LMXB leads to the formation of common envelope. After common envelope phase the primary core rapidly evolves towards core collapse and the BH formation. The hight mass x-ray binary system evolved as a detached system. The difference in distribution of the masses of black holes in LMXB and HMXB is shown in Fig. \ref{ris:Distrb}. The typical black hole masses in LMXB are around $M_{BH} \simeq 8 M_{\odot}$. The  black hole mass in HMXB is distributed in broad range of masses without any peak. 

The standard evolutionary scenarios require the donor mass distribution peaks to be, at least, $\sim 1\; M_{\odot}$ to eject a massive envelope of the black hole progenitor \citep{Podsiadlowski_2010}. The low mass companion with mass below $ 1\; M_{\odot}$ has difficulties in ejecting the tightly bound envelope of the massive primary during the spiral-in process. But observational masses in BH LMXB are found in the mass range 0.1 - 1.6 $M_{\odot}$ with peak at 0.6 $M_{\odot}$. \citet{Wang_2016} use both stellar evolution and binary population synthesis to study the evolutionary history of BH LMXBs. It has been shown that it is possible to form BH LMXBs with optical components masses below  $\sim 1\; M_{\odot}$ and the standard CE scenario if most BHs are born through failed supernovae. But $\alpha$-elements (O, Si, Mg, etc.), which cannot be produced by nuclear burning in low mass stars were detected in the atmosphere of two LMXB optical components (GRO J1655-40 and SAX J1819.3-2525). Therefore, the predictions are still in tension with available observations.

The mass of black hole in LMXB is given by:

\begin{eqnarray} \label{eq:massfun}
M_{x} &=& \frac{f_{v}(M)\left(1+q^{-1}\right)^{2}}{\sin^{3} i},
\end{eqnarray} 
where $q=M_x/M_v$, $M_x$ - the mass of the X-ray component, $i$ - the orbital inclination, $f_{v}(M)$ - the mass function of the optical component. 
The mass function is derived from:
\begin{eqnarray} \label{eq:defnussfunc}
f_v(M)=1.038\cdot 10^{-7} K_c^3P_{orb}(1-e^2)^{3/2}.
\end{eqnarray}

where $K_c$ - the semi-amplitude of the radial velocity curve of the center of mass of the companion star, $P_{orb}$ - the orbital period, $e$ - the eccentricity. In case of stellar mass black holes ($q=M_x/M_v \gg 1$) mass $M_x$ depends weakly on $q$ (see \ref{eq:massfun} ), so we can consider the black hole mass in  BH LMXB as a function of the $K_c$ and $i$. However for the mass of the optical star $ M_v = M_x / q $ value of $q$ is significant.

The effects of the asphericity of a star in particular BH LMXB on its rotational broadening of absorption lines was considered by Marsh et al. \citep{Marsh1994} and Shahbaz \citep{Shahbaz2003}. Marsh et al. \citep{Marsh1994} employed both the geometry of Roche lobe and orbital smearing of spectral lines to estimate the mass ratio systematic uncertainty in BH LMXB A0620-00. This uncertainty was about 5\%. It needs to be noted that their findings were obtained without full model-atmosphere calculations.

Shahbaz \citep{Shahbaz2003} found mass ratio $q$ in Nova Sco 1994 (the F star) fitting the observed high quality spectrum with synthetic spectra. Author made direct use of NextGen atmosphere models  which take into account the varying temperature and gravity across the secondary star’s photosphere, by incorporating the synthetic spectra into the secondary star’s Roche geometry. This method of determination of the mass ratio does not depend on assumptions about the rotation profile and limb darkening coefficients. Also the author performed the calculations  of mass ratio with star's model as disk with uniform local profile and linear limb darkening law so-called classical rotational broadening model \citep{Collins_1995}. The calculations for two different values of limb-darkening coefficients  showed that the Roche-model gives a mass ratio that lies in between the values obtained using limb-darkening coefficients of zero and 0.52 (continuum).  

Antokhina and Cherepashchuk \citep{Pisma_1997} also investigated tidal distortion effect on mass ratio determination using rotational broadening of absorption lines. They showed that neglecting the pear-like shape of the star leads to underestimation of the component-mass ratio $q = M_x/M_v$, i.e., to overestimation of the mass of the black hole in the system.

In this paper, the accuracy of two widely used approximations of optical star in LMXB mass ratio estimation (spherical model and classical rotational broadening model)  has been compared.

\section{The method of mass ratio estimation by rotational broadening of absorption lines}

If the orbit is circular and the star’s rotation is synchronized with the orbital motion, the following kinematic relation is satisfied \citep{AnMihKniga,Wade_Horne_1988}:
\begin{eqnarray}\label{eq:Wade_Horne}
V_{rot}\sin i=K_{c}\left(1+\frac{1}{q}\right)\frac{r_v}{a}.
\end{eqnarray}

Here, $q = M_x/M_v$ is the ratio of the masses of the relativistic object and the optical star in the binary, $K_c$ the semi-amplitude of the radial velocity curve for the center of mass of optical star, $r_v$ is mean star radius, $a$ is the radius of the circular orbit, $V_{rot}$ is the equatorial
rotational velocity, and $i$ is the orbital inclination.

Paczynski \citep{Pach_1976} proposed an approximate analytical expression for calculating the radius of a sphere with a volume equal to one of the critical Roche lobe:
\begin{eqnarray} \label{eq:Pach_radius}
\frac{\bar R}{a}\simeq 0.462\; \left( 1+q\right)^{-1/3}.
\end{eqnarray}

The  $\bar R$ value is usually called the “mean radius of
the Roche lobe.” Formulas (\ref{eq:Wade_Horne}) and (\ref{eq:Pach_radius}) can be used
to derive the formula that was first presented by Wade and Horne  \citep{Wade_Horne_1988}:

\begin{eqnarray} \label{eq:Pach}
\frac{V_{rot}\sin i}{K_{c}}\simeq 0.462\; q^{-1/3}\left( 1+\frac{1}{q}\right)^{2/3}.
\end{eqnarray}

The expression (\ref{eq:Pach}) is widely used to determine the component-mass ratios $q$ in binary systems based on the rotational broadening of lines (see, e.g. \cite{Marsh1994, Casares1994}). The more exact formula of Eggleton \citep{Eggleton_1983} can also be used for the mean radius of the Roche lobe, but formula (\ref{eq:Pach}) will be used further in the calculations. It is necessary to mention that in the limit case where $q=M_x/M_v >> 1$ (this case corresponds to BH LMXB) mass ratio $q$ is given by:
\begin{eqnarray} \label{eq:big_q}
q \simeq \left( \frac{0.462 K_c}{V_{rot} \sin i } \right)^{3}.
\end{eqnarray}
It follows from the expression (\ref{eq:big_q})  that small uncertainties in $V_{rot} \sin i$  lead to large uncertainties in $q$.

Marsh et al. \citep{Marsh1994} and Casares \& Charles \citep{Casares1994} determined the rotational broadening of line profiles in the spectra of the optical components of X-ray binary systems. They used the observed line profiles for slowly rotating stars with close spectral types (reference stars) as profiles unbroadened by rotation. The spectra of the stars in the X-ray binaries and those of the single slowly rotating stars were obtained with the same spectral resolution. The rotational broadening of the spectra of the reference stars was modelled, assuming that these profiles would be the same in the absence of rotation, identifying the value of $V_{rot} \sin i$ for which the spectra of the single star and the star in the X-ray binary were in best agreement with $\chi^2$ criterion.  
In this way, the rotational velocity $V_{rot} \sin i$ of the star in the binary system was found. This approach made it possible not to correct the influence of the instrumental function of the spectrograph. 

Two methods, i.e the minimum value of $\chi^2$ and equality of Full Width Half Maximum ($FWHM$) for  CaI $\lambda 6439.075 \; \angstrom $  line profiles, have been employed to estimate $V_{rot} \sin i $. In our current study, we computed both the minimum value of $\chi^2$ and equality of Full Width Half Maximum ($FWHM$) for  CaI $\lambda 6439.075 \; \angstrom $  line profiles. 
We found that difference in determination of $V_{rot} \sin i $ by these two methods do not exceed  $2 \%$. Therefore, we use $FWHM$ equality method.

\section{The spherical star approximation}

We used the computational method proposed by Antokhina et al. \citep{Antokhina_Shim_2005} to correctly model theoretically the line profiles emitted by optical stars in close X-ray binary systems. It is necessary to describe this algorithm briefly.
The computation of the theoretical line profiles and radial velocity curves is carried out using the
synthesis method. The tidally deformed stellar surface in the Roche model is divided into thousands of area elements. The flux of emergent radiation is calculated for each area, taking into account gravitational darkening, heating of the surface by incident radiation from the companion (the reflection effect), and limb darkening. The local line profile for an area element is calculated by constructing an atmosphere model  for the given point of the stellar surface. The spectrum
of the external radiation from the compact source is specified using X-ray observations or a model function.

Thus, an atmosphere model is calculated at each point of the stellar surface, for specified values of $T_{loc}$, $g_{loc}$, and $k^{loc}_x$, by solving the radiative transfer equation
in the line in the presence of the incident external X-ray flux \citep{Antokhina_Shim_2005}. The adopted model atmosphere is then used to calculate the intensity of the emergent line and continuum radiation, i.e., the line profile for each area element. At each orbital phase, the line
is summed over the visible surface of the star, taking into account the Doppler effect, and the integrated line profile from the stellar disk is constructed.

We calculated  the ``accurate'' theoretical CaI $\lambda 6439.075 \; \angstrom $ profiles and the radial velocity curves using the code  \cite{Antokhina_Shim_2005} for fixed set of parameters: masses $M_x, M_v$, Roche lobe filling factor $\mu =1$, the effective temperature of the optical star  $T_{eff}$, the orbital inclination $i$, e.t.c. at 20 different orbital phases.

The projected rotational velocity of Roche lobe filling star $V_{rot}^{Roche} \sin i$ is one of the output parameter of our calculation. Note that $V_{rot}^{Roche} \sin i$ in current orbital phase is a projected rotational velocity of spherical star with equal volume. We determined $FWHM$ of ``true'' theoretical profiles in each orbital phase and average over orbital period (phase
$\phi=0$ corresponds to the eclipse of the X-ray source by the optical star). The shape and $FWHM$ of the profiles depend on orbital phase since the projection of the Roche lobe filling star onto the plane of the sky has various areas \citep{Pisma_1997}.  

We estimated the rotational broadening and mass ratio $q$ by fitting ``true'' theoretical profiles with the spherical star approximation to obtain accuracy of mass ratio determination.
Such mass ratio will be denoted as $q_{sph}$. 
 
The results of modelling using code \cite{Antokhina_Shim_2005} are listed in Table \ref{tabular:ParametersP1i90} - \ref{tabular:ParametersP4i45}. We consider the optical star
with mass $M_v=0.8 \; M_{\odot}$ and the effective temperature $T_{eff}=5000 \; K$. The Roche lobe filling factor of the stars is taken to be $\mu=1$. The orbital inclination is set to be $i=90^o$. To illustrate
the effect of the inclination angle we have also computed the parameters of LMXB for $i=45^o$ (see table \ref{tabular:ParametersP4i45}). The effect of X-ray heating is ignored $k_x=L_x/L_{v}^{bol}=0$. The mass ratio is in the range of $q=1-40$, and the period is taken to be $P_{orb}=1,2,4$ d.

In the table \ref{tabular:ParametersP1i90} $\bar R$ is the mean radius of the Roche lobe, $\bar g$ is the mean surface gravity, $a$ is the radius of the circular orbit, $K_v$ is is the semi-amplitude of the
stellar radial velocity curve in the Roche model, $K_c$ is the semi-amplitude of the radial velocity curve of the stellar barycenter. It is important to emphasize that the mean radius of the Roche lobe $\bar R$ was defined by integration over the Roche lobe volume in Table \ref{tabular:ParametersP1i90} - \ref{tabular:ParametersP4i45}. Parameter $V_{rot}$ is the accurate value rotational velocity of Roche lobe filling star.  

It is shown in Table \ref{tabular:ParametersP1i90} - \ref{tabular:ParametersP4i45} that varying the mass ratio with fixed period  $P_{orb}$ leads to the change of the radius of the the circular orbit $a$, accordingly, to the change of the semi-amplitude radial velocity curve $K_c$ 

The variation of mean radius and rotational velocity is very small. In this way mass ratio $q$ is a function of semi-amplitude of the stellar radial velocity curve $K_c$. We fixed the radius of the circular orbit $a$, varied $P_{orb}$ and list the parameters in Table \ref{tabular:Rfix} 

We introduce a Cartesian coordinate system ($X$, $Y$,$Z$) with its origin at the center of mass of
the optical star. The $X$ axis is directed along the line passing through the component centers, the
$Y$ axis lies in the orbital plane, and the $Z$ axis is perpendicular to the orbital plane (see \cite{Antokhina_1994,Pisma_1997}). Then $r_{point}$ and $r_{back}$ are the points of intersections of the Roche lobe with  $OX$ axis. Distances between center of mass of the optical star and of intersections of the Roche lobe with  $OY$ and $OZ$ axes are $r_{side}$ and $r_{pole}$ respectively. We list the geometrical parameters of the optical star for $q=1-40$ in Table \ref{tabular:Roche}.
The elongation along the centers of mass is $r_{point}/r_{pole}$ is also shown. 

The goal of our study is determination of rotational broadening of theoretical CaI $\lambda 6439.075 \; \angstrom $ profiles using two models of optical star. We consider the spherical star model in the first place. Using code \cite{Antokhina_Shim_2005}, we iterated $V_{rot} \sin i$ for spherical star model at which Roche lobe filling and spherical star's line profiles have the same $FWHM$.  If the star is spherical the line profile is independent of the orbital inclination $i$ and orbital phase $\phi$. However, such a dependence arises for a star that fills its Roche lobe and has a tidally deformed (pear-like) shape.

We list in table \ref{tabular:Sphere} results of our calculation of $V_{rot}$ for the spherical star model. We consider the optical star with mass $M_v=0.8 \; M_{\odot}$ and the effective temperature $T_{eff}=5000 \; K$. The Roche lobe filling factor of the stars is taken to be $\mu=1$. The orbital inclination is set to be $i=90^o, \; 45^o, \; 60^o,\; 70^o$ and $90^o$. The effect of X-ray heating is ignored $k_x=L_x/L_{v}^{bol}=0$. The mass ratio is in the range of $q=1-40$, and the period is taken to be $P_{orb}=2$ d. The mass ratio  $q_{sph}$ founded by the formula (\ref{eq:Pach}) also given in table \ref{tabular:Sphere}.

\begin{figure}[H]
\center{\includegraphics[width=140mm,scale=0.8]{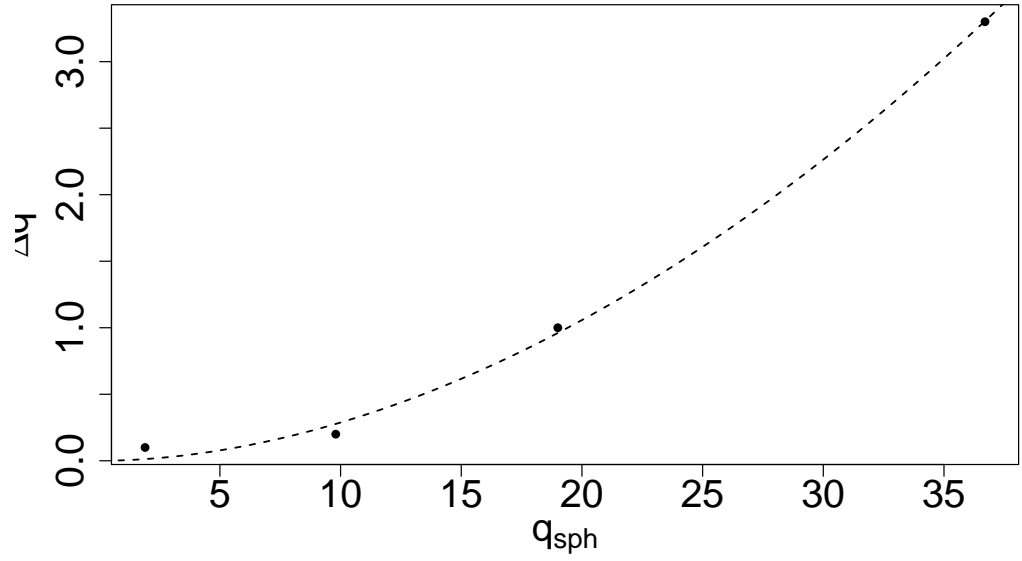}}\\
\caption{$\Delta q $ correction as a function of mass ratio $q_{sph}$ in case of the spherical star approximation. Corrected $q$ is expressed by formula $q=q_{sph}+\Delta q$.}
\label{ris:Sphere}
\end{figure}

The line is maximally broad (the $FWHM$ is maximum) in the absence of X-ray heating ($k_x = 0$) at quadrature (phase $\phi=0.25$) since the projection of the star onto the plane of the sky has its maximum area. Therefore, the spherical star model has the highest value of $V_{rot} \sin i$ this orbital phase.
Hence, the value of $ q_{sph} $ is less than $ q $ according the equation (\ref{eq:Pach}). For the case of $\phi=0$ (phase $\phi=0$ corresponds to the eclipse of the X-ray source by the optical star) the projection of the star onto the plane of the sky has its minimum and the value of $ q_{sph} $ is higher than $ q $. The  phase-averaged values of  $q_{sph}$ is less then ``true'' theoretical  $q=M_x/M_v$. The deviation between $q_{sph}$ and $ q $ is small (is lower then $5\%$) in the case of small value of $q=M_x/M_v$. However, the deviation reaches $10\%$ in the case of high value of $q=M_x/M_v$ due to a strongly nonlinearity of the equation (\ref{eq:Pach}). The difference between $q_{sph}$ and $q=M_x/M_v$ is
$$
\Delta q = q -  q_{sph}
$$
The Fig \ref{ris:Sphere} shows how the mass ratio $q_{sph}$ estimated from spherical star model starts to deviate from ``accurate'' theoretical  $q=M_x/M_v$.

\section{The classical rotational broadening model}

A rougher method is commonly used (see e.g. \cite{Marsh1994, Casares1994}). 
Let a line with central wavelength $\lambda_0$ have a profile $I_{loc}(\lambda - \lambda_0)$, and not be subject to Doppler broadening, with the line profile being constant
over the stellar disk. The profile of a line emitted by a star rotating with velocity $V_{rot} \sin i$ is given by \citep{Shajn1929,Carrol1933,Collins_1995}:
\begin{eqnarray} \label{eq:Rot_conv}
I_{Dop}(\lambda-\lambda_{0})= \int_{-1}^{+1} I_{loc}\left(\lambda-\lambda_{0}-\lambda_{0} x\; \frac{V_{rot}}{c} \sin i \right)\cdot A(x)dx.
\end{eqnarray}
Here, $i$ is the orbital inclination, $c$ is the speed of light, and $A(x)$ is the rotational broadening function, which indicates how the profile of an infinitely narrow absorption line in the stellar spectrum is transformed by the Doppler effect due to rotation.

We take the limb darkening of the stellar disk to be described by the linear law:
\begin{eqnarray} \label{eq:Limb_darkening}
I(\vartheta) = I(0)(1-u+u \cos\vartheta ),
\end{eqnarray} 
where $u$ is the linear limb-darkening coefficient, $\vartheta$ is  the angle between the normal to the stellar surface at a given point and the direction toward the observer, and $I(0)$  is the intensity emitted per unit area at the center of the stellar disk. The function $A(x)$ then has the form (see e.g. \cite{Gray_2005}):
\begin{eqnarray}\label{eq:A}
A(x) =C_1 \cdot\sqrt{(1-x)^2}+C_2\cdot(1-x^2).
\end{eqnarray}
where

\begin{eqnarray}\label{eq:c1}
C_1=\frac {2(1-u)}{\pi(1-u/3)}.
\end{eqnarray}

\begin{eqnarray}\label{eq:c2}
C_2 =\frac{u}{2(1-u/3)}.
\end{eqnarray}

\begin{figure}[H]
\center{\includegraphics[width=140mm,scale=0.8]{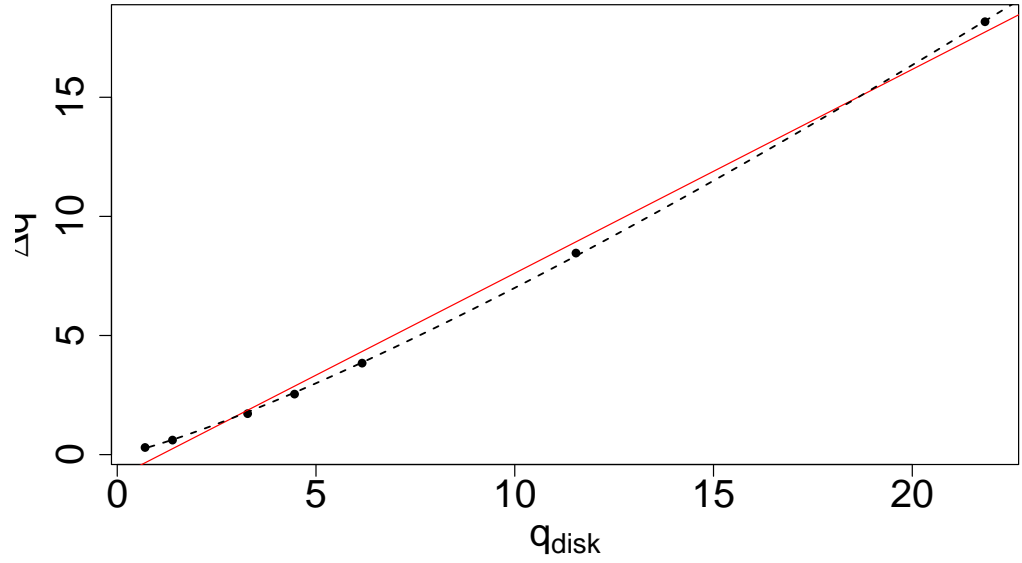}}\\
\caption{$\Delta q $ correction as a function of mass ratio $q_{disk}$ in case of the classical rotational broadening model. Solid line represent the linear fitting. A more realistic non-linear fitting represented by a dashed line. Corrected $q$ is expressed by formula $q=q_{disk}+\Delta q$.}
\label{ris:deltaq}
\end{figure}

Following the prescription of previous chapter we found $V_{rot}^{disk} \sin i$ and $q_{disk}$ have the same $FWHM$ for classical rotational broadening model at which Roche lobe filling star's  line profile and the line profile of a template star, had been convolved with the standard rotation profile, .  The accuracy of the method is the difference between derived value $q_{disk}$ and fixed ``true'' value $ q $. We use theoretical CaI $\lambda 6439.075 \; \angstrom $ profile for orbital inclination $i=0^o$, $T_{eff}=5000 \; K$ and Roche lobe filling factor $\mu = 0.5$ as an unbroadened profile in the spectra of non-rotating template star. The linear limb-darkening coefficient $u=0.66$ was assumed in deriving the rotational broadening, based on the work of Al-Naimiy \citep{Al_Naimiy1977}. To perform convolution with the standard rotation profile  we used PyAstronomy 0.9 software (\texttt{rotBroad} function).
Table \ref{tabular:Disk} lists the results of $V_{rot}^{disk}$ and $q_{disk}$ calculations. We also added the `true'' values of  $V_{rot}^{Roche}$ for comparison with $V_{rot}^{disk}$ in table \ref{tabular:Disk}. It can be seen that $V_{rot}^{disk}$ is completely higher than the $V_{rot}^{Roche}$. The deviation between $V_{rot}^{disk}$ and  $V_{rot}^{Roche}$ can reach $\sim 20\%$ at high value of $q=M_x/M_v$. These estimates $V_{rot}^{disk}$ correspond to the $q_{disk}$ that is less than $q$ (see eq. (\ref{eq:Pach})). We have calculated the $\Delta q$ correction for both linear (dashed line) and non linear fitting (solid line) as it is shown in Fig. \ref{ris:deltaq}. Hence, for a given mass ratio $q_{disk}$ the correction $\Delta q$ is provided by:
\begin{eqnarray}\label{eq:NonLinFitq}
\Delta q = ( 0.41 \pm 0.01){q_{disk}}^{1.224 \pm 0.008},
\end{eqnarray} 
or the more rough linear fitting:
\begin{eqnarray}\label{eq:LinFitq}
\Delta q = (0.85 \pm 0.03)q_{disk} - (0.94 \pm 0.28)
\end{eqnarray}

As a second step we explore the effect of $T_{eff}$ variation in the accuracy of $q_{disk}$. Our calculation is showed that equations (\ref{eq:NonLinFitq} - \ref{eq:LinFitq}) are accurate to better than  $5 \; \%$ in effective temperature range $ 4000$ - $8000 \; K$, and better than  $1.5 \; \%$ in effective temperature range $ 5000$ - $7000 \; K$. The limb-darkening coefficient varied in accordance with the effective temperature of the star $T_{eff}$ \citep{Al_Naimiy1977}.

\begin{figure}[H]
\center{\includegraphics[width=140mm,scale=0.8]{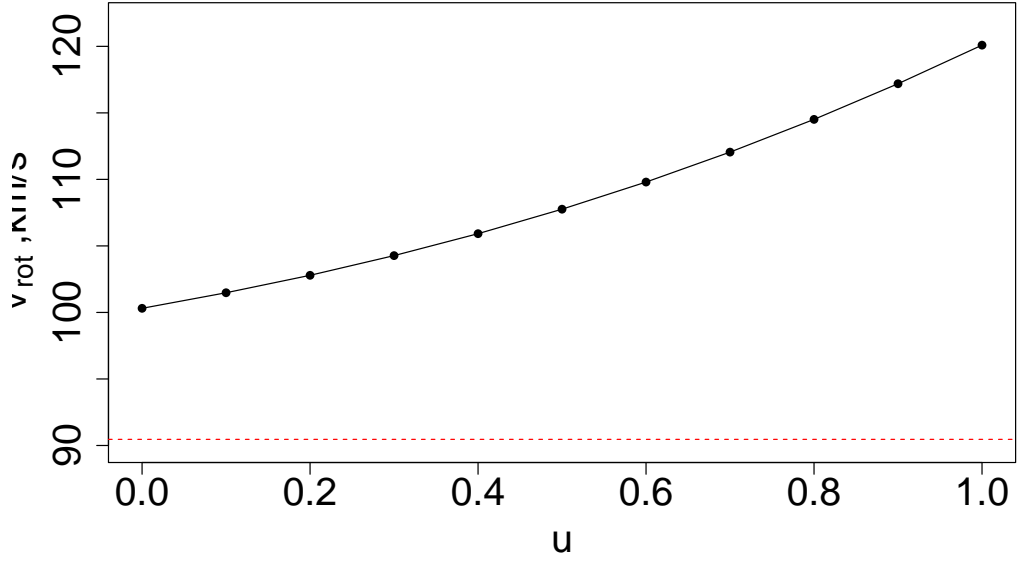}}\\
\caption{The projected equatorial rotational velocity $V_{rot}^{disk} \sin i$ restored in the framework of classical rotational broadening model with the linear limb-darkening coefficient $u$. Note that the $FWHM$ of the ``observed'' line profiles have been fixed ($FWHM = 3.685 \; \angstrom$). Dashed line represent the $V_{rot} = 90.41$ km/s obtained in a Roche model.}
\label{ris:var_u}
\end{figure}

The linear limb-darkening coefficient that we have used in eq. (\ref{eq:Rot_conv}) is appropriate to the continuum and may not apply to CaI $\lambda 6439.075 \; \angstrom $ absorption line \citep{Collins_1995}. We try choose limb-darkening coefficient to offset $V_{rot}^{disk}$. We have computed  the theoretical CaI $\lambda 6439.075 \; \angstrom $ profiles in a Roche model using the code  \cite{Antokhina_Shim_2005} for fixed set of parameters $q=M_x/M_v=10$, $i=90^o$, $P=1$ d  and obtain $FWHM_{mean}=3.685\; \angstrom$ and $V_{rot}=90.46$ km/s (see table \ref{tabular:ParametersP1i90}). This value is plotted in Fig. \ref{ris:var_u} in dashed line. From Fig. \ref{ris:var_u} it can be seen that there are no values of the linear limb-darkening coefficient $u$ in the range from 0 to 1 that allow to offset $V_{rot}^{disk}$ for CaI $\lambda 6439.075 \; \angstrom $ absorption line if we fixed $FWHM$.

\section{Conclusions}

We compare the accuracy of the mass ratio $q$ determination using two models of optical star in BH LMXB: the spherical star model and the classical rotational broadening model. We use the theoretical CaI $\lambda 6439.075 \; \angstrom $ profiles in the Roche model computed with code \cite{Antokhina_Shim_2005} as ``observed'' line profiles. As in the case of spherical star the integrated line profile is constructed by solving the radiative transfer equation like in the case the Roche model of optical star.

\begin{figure}[H]
\center{\includegraphics[width=140mm,scale=0.8]{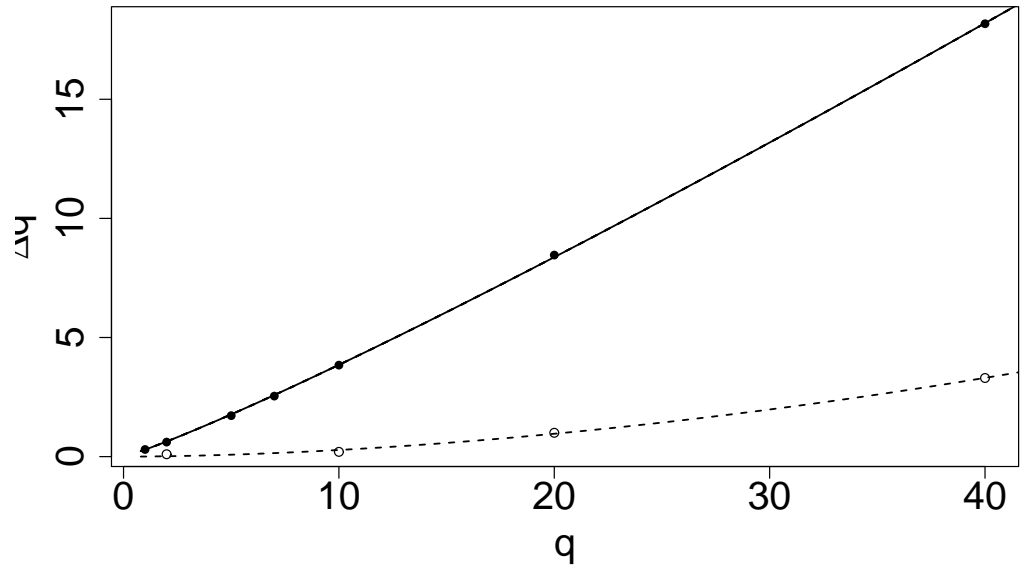}}\\
\caption{ Correction  $\Delta q $ as a function of $q$ (modelled value) for the spherical star model (dashed) and the classical rotational broadening model (solid).}
\label{ris:two_deltas}
\end{figure}

As a result, we determine the corrections $\Delta q$ to mass ratio $q=M_x/M_v$. In both cases the mass ratios estimated from star models are lower than ``true'' theoretical  $q=M_x/M_v$. But in the case of spherical star model the deviation between $q_{sph}$ and $ q $ reaches $10\%$ only. Hoverer, in the case of the classical rotational broadening model the deviation between $q_{disk}$ and $q=M_x/M_v$ can reach 1.5 times at high value of $q=M_x/M_v  \geq 30$ (see Fig. \ref{ris:two_deltas}). Note, there are no values of the linear limb-darkening coefficient $u$ in the range from 0 to 1 that allow to offset $q_{disk}$ for  CaI $\lambda 6439.075 \; \angstrom $ absorption line.  

\bibliographystyle{apj}
\bibliography{main}

\newpage

\begin{table}[H]
\caption{Input and derived parameters for $P_{orb} =1$ d and $i=90^o$ (the mass ratio varies in the range $q=1-40$)}
\label{tabular:ParametersP1i90}
\begin{center}
\begin{tabular}{lccccccc}
\hline
\hline
Parameter&$q=1$&$q=2$&$q=5$&$q=7$&$q=10$&$q=20$&$q=40$\\
\hline
\hline
$M_v\;(M_{\odot})$&0.8&0.8&0.8&0.8&0.8&0.8&0.8\\
$M_x\;(M_{\odot})$&0.8&1.6&4.0&5.6&8.0&16.0&32.0\\
$P_{orb}\;(d)$&1&1&1&1&1&1&1\\
$a\;(R_{\odot})$&4.93&5.64&7.11&7.82&8.70&10.79&13.49\\
$\log \bar g$&3.762&3.795&3.808&3.808&3.805&3.796&3.786\\
$\bar R\;(R_{\odot})$&1.87&1.81&1.78&1.78&1.79&1.80&1.82\\
$K_v\;(km/s)$&123.58&189.03&298.45&345.10&398.84&518.85&664.93\\
$K_c\;(km/s)$&124.72&190.36&299.81&346.48&400.29&520.22&666.05\\
$V_{rot}\;(km/s)$&94.77&91.58&90.19&90.23&90.46&91.24&92.18\\
\multicolumn{8}{c}{  CaI $\lambda 6439.075 \; \angstrom $ line}\\
$FWHM_{\phi=0.0}\;(\angstrom)$&3.780&3.629&3.554&3.549&3.552&3.569&3.594\\
$FWHM_{\phi=0.25}\;(\angstrom)$&3.940&3.829&3.792&3.804&3.823&3.864&3.903\\
$FWHM_{mean}\;(\angstrom)$&3.863&3.729&3.673&3.675&3.685&3.717&3.755\\
\hline
\end{tabular}
\end{center}\end{table}

\begin{table}[H]
\caption{Input and derived parameters for Roche model (the mass ratio varies in the range $q=1-40$, $P_{orb} =2$ d and $i=90^o$)}
\label{tabular:ParametersP2i90}
\begin{center}
\begin{tabular}{lccccccc}
\hline
\hline
Parameter&$q=1$&$q=2$&$q=5$&$q=7$&$q=10$&$q=20$&$q=40$\\
\hline
\hline
$M_v\;(M_{\odot})$&0.8&0.8&0.8&0.8&0.8&0.8&0.8\\
$M_x\;(M_{\odot})$&0.8&1.6&4.0&5.6&8.0&16.0&32.0\\
$P_{orb}\;(d)$&2&2&2&2&2&2&2\\
$a\;(R_{\odot})$&7.82&8.96&11.28&12.42&13.81&17.13&21.41\\
$\log \bar g$&3.336&3.393&3.407&3.406&3.404&3.395&3.385\\
$\bar R\;(R_{\odot})$&2.97&2.87&2.83&2.83&2.84&2.86&2.89\\
$K_v\;(km/s)$&98.03&150.04&236.82&273.89&316.63&411.84&527.65\\
$K_c\;(km/s)$&98.99&151.09&237.96&275.00&317.71&412.90&528.65\\
$V_{rot}\;(km/s)$&75.22&72.69&71.59&71.61&71.80&72.42&73.16\\
\multicolumn{8}{c}{ CaI $\lambda 6439.075 \; \angstrom $ line}\\
$FWHM_{\phi=0.0}\;(\angstrom)$&3.003&2.892&2.819&2.815&2.817&2.833&2.857\\
$FWHM_{\phi=0.25}\;(\angstrom)$&3.166&3.079&3.049&3.057&3.072&3.113&3.154\\
$FWHM_{mean}\;(\angstrom)$&3.078&2.972&2.927&2.928&2.936&2.996&2.996\\

\hline
\end{tabular}
\end{center}\end{table}

\begin{table}[H]
\caption{Input and derived parameters for Roche model (the mass ratio varies in the range  $q=1-40$,  $P_{orb} =4$ d and $i=90^o$)}
\label{tabular:ParametersP4i90}
\begin{center}
\begin{tabular}{lccccccc}
\hline
\hline
Parameter&$q=1$&$q=2$&$q=5$&$q=7$&$q=10$&$q=20$&$q=40$\\
\hline
\hline
$M_v\;(M_{\odot})$&0.8&0.8&0.8&0.8&0.8&0.8&0.8\\
$M_x\;(M_{\odot})$&0.8&1.6&4.0&5.6&8.0&16.0&32.0\\
$P_{orb}\;(d)$&4&4&4&4&4&4&4\\
$a\;(R_{\odot})$&12.42&14.22&17.91&19.71&21.92&27.19&33.99\\
$\log \bar g$&2.959&2.992&3.01&3.005&3.002&2.994&2.983\\
$\bar R\;(R_{\odot})$&4.72&4.56&4.49&4.49&4.50&4.54&4.59\\
$K_v\;(km/s)$&77.77&119.03&187.98&217.38&251.27&326.86&418.84\\
$K_c\;(km/s)$&78.57&119.92&188.87&218.27&252.17&327.72&419.59\\
$V_{rot}\;(km/s)$&59.70&57.69&56.82&56.84&56.99&57.48&58.07\\
\multicolumn{8}{c}{ CaI $\lambda 6439.075 \; \angstrom $ line}\\
$FWHM_{\phi=0.0}\;(\angstrom)$&2.380&2.290&2.233&2.230&2.231&2.244&2.264\\
$FWHM_{\phi=0.25}\;(\angstrom)$&2.521&2.443&2.423&2.429&2.442&2.476&2.509\\
$FWHM_{mean}\;(\angstrom)$&2.443&2.359&2.323&2.324&2.331&2.353&2.380\\
\hline
\end{tabular}
\end{center}\end{table}

\begin{table}[H]
\caption{Input and derived parameters for Roche model (the mass ratio varies in the range $q=1-40$, $P_{orb} =4$ d and $i=45^o$)}
\label{tabular:ParametersP4i45}
\begin{center}
\begin{tabular}{lccccccc}
\hline
\hline
Parameter&$q=1$&$q=2$&$q=5$&$q=7$&$q=10$&$q=20$&$q=40$\\
\hline
\hline
$M_v\;(M_{\odot})$&0.8&0.8&0.8&0.8&0.8&0.8&0.8\\
$M_x\;(M_{\odot})$&0.8&1.6&4.0&5.6&8.0&16.0&32.0\\
$P_{orb}\;(d)$&4&4&4&4&4&4&4\\
$a\;(R_{\odot})$&12.42&14.22&17.91&19.71&21.92&27.19&33.99\\
$\log \bar g$&2.959&2.992&3.01&3.005&3.002&2.994&2.983\\
$\bar R\;(R_{\odot})$&4.72&4.56&4.49&4.49&4.50&4.54&4.59\\
$K_v\;(km/s)$&54.86&84.11&132.85&153.66&177.66&231.02&296.01\\
$K_c\;(km/s)$&55.56&84.80&133.55&154.34&178.31&231.73&296.69\\
$V_{rot}(km/s)$&59.70&57.69&56.82&56.84&56.99&57.48&58.07\\
\multicolumn{8}{c}{  CaI $\lambda 6439.075 \; \angstrom $ line}\\
$FWHM_{\phi=0.0}\;(\angstrom)$&1.646&1.580&1.543&1.542&1.543&1.553&1.566\\
$FWHM_{\phi=0.25}\;(\angstrom)$&1.770&1.722&1.708&1.715&1.723&1.744&1.766\\
$FWHM_{mean}\;(\angstrom)$&1.718&1.658&1.635&1.636&1.642&1.658&1.676\\

\hline
\end{tabular}
\end{center}\end{table}

\begin{table}[H]
\caption{Input and derived parameters for $P_{orb} =4$ d and $i=45^o$ ($q=1-40$, $a=12.42 \; R_{\odot}$ and $i=90^o$)}
\label{tabular:Rfix}
\begin{center}
\begin{tabular}{lccccccc}
\hline
\hline
Parameter&$q=1$&$q=2$&$q=5$&$q=7$&$q=10$&$q=20$&$q=40$\\
\hline
\hline
$M_v\;(M_{\odot})$&0.8&0.8&0.8&0.8&0.8&0.8&0.8\\
$M_x\;(M_{\odot})$&0.8&1.6&4.0&5.6&8.0&16.0&32.0\\
$P_{orb}\;(d)$&4.0&3.266&2.310&2.0&1.705&1.234&0.883\\
$a\;(R_{\odot})$&12.42&12.42&12.42&12.42&12.42&12.42&12.42\\
$\log \bar g$&2.959&3.324&3.324&3.406&3.496&3.675&3.858\\
$\bar R\;(R_{\odot})$&4.72&3.98&3.114&2.83&2.55&2.07&1.677\\
$K_v\;(km/s)$&77.77&127.4&225.81&273.89&333.96&483.72&692.92\\
$K_c\;(km/s)$&78.57&128.31&226.8&275.0&335.07&485.01&694.15\\
$V_{rot}(km/s)$&59.70&61.72&68.23&71.61&75.72&85.06&96.06\\
\multicolumn{8}{c}{  CaI $\lambda 6439.075 \; \angstrom $ line}\\
$FWHM_{\phi=0.0}\;(\angstrom)$&2.380&2.444&2.694&2.815&2.974&3.334&3.750\\
$FWHM_{\phi=0.25}\;(\angstrom)$&2.521&2.611&2.911&3.057&3.234&3.618&4.075\\
$FWHM_{mean}\;(\angstrom)$&2.443&2.524&2.792&2.928&3.095&3.472&3.912\\

\hline
\end{tabular}
\end{center}\end{table}

\begin{table}[H]
\caption{ The relative radii of a star filling its Roche lobe for $q=1-40$}
\label{tabular:Roche}
\begin{center}
\begin{tabular}{lccccccc}
\hline
\hline
parameter&$q=1$&$q=2$&$q=5$&$q=7$&$q=10$&$q=20$&$q=40$\\
\hline
\hline
$r_{point}/a$&0.5000&0.4291&0.3414&0.3119&0.2825&0.2313&0.1877\\
$r_{side}/a$&0.3740&0.3129&0.2422&0.2195&0.1974&0.1598&0.1287\\
$r_{pole}/a$&0.3561&0.2998&0.2329&0.2112&0.1899&0.1536&0.1237\\
$r_{back}/a$&0.4050&0.3454&0.2746&0.2513&0.2283&0.1884&0.1547\\
$r_{point}/r_{pole}$&1.404&1.431&1.468&1.477&1.488&1.506&1.517\\
\hline
\end{tabular}
\end{center}\end{table}

\begin{table}[H]
\small
\caption{Approximate values $V_{rot}$ and $q_{sph}$, with Ca I $\lambda 6439 \; \angstrom $ line for spherical star model ($P_{orb}=2$ d).}
\label{tabular:Sphere}
\begin{center}
\begin{tabular}{c|c||c|c|c||c|c|c||c|c|c}
\hline
\multirow{2}{*}{i}&\multirow{2}{*}{$K_v $ (km/s)}&\multicolumn{3}{c||}{$\phi=0.0$}&\multicolumn{3}{c||}{$\phi=0.25$}&\multicolumn{3}{c}{the phase-averaged values}\\
\cline{3-11}
&&\footnotesize{FWHM,\angstrom}&$V_{rot}$&$q_{sph}$&\footnotesize{FWHM,\angstrom}&$V_{rot}$&$q_{sph}$&\footnotesize{FWHM,\angstrom}&$V_{rot}$&$q_{sph}$\\
\hline
\multicolumn{11}{c}{\textbf{q=2}}\\
\hline
30$^o$& 74.8&1.386&70.0&2.1&1.530&77.0&1.8&1.474&74.4&1.9\\
45$^o$&105.9&1.996&71.0&2.1&2.169&77.0&1.8&2.095&74.2&1.9\\
60$^o$&129.8&2.470&71.3&2.0&2.658&76.6&1.8&2.573&74.3&1.9\\
70$^o$&141.0&2.688&71.3&2.0&2.892&76.5&1.8&2.794&74.3&1.9\\
90$^o$&150.0&2.892&71.9&2.0&3.079&76.5&1.8&2.972&74.1&1.9\\

\hline
\multicolumn{11}{c}{\textbf{q=10}}\\
\hline

30$^o$&158.1&1.360&68.6&11.4&1.531&77.0&8.5&1.459&73.7&9.5\\
45$^o$&223.8&1.949&69.2&11.2&2.168&76.9&8.6&2.072&73.5&9.6\\
60$^o$&274.1&2.413&69.8&11.0&2.656&76.6&8.6&2.543&73.3&9.7\\
70$^o$&141.0&2.688&71.3&10.9&2.892&76.5&8.7&2.794&74.3&9.7\\
90$^o$&316.6&2.817&70.3&10.8&3.072&76.4&8.7&2.937&73.1&9.8\\

\hline
\multicolumn{11}{c}{\textbf{q=20}}\\
\hline

30$^o$&205.7&1.366&69.0&22.9&1.550&78.1&16.3&1.474&74.4&18.8\\
45$^o$&291.0&1.961&69.7&22.9&2.195&77.9&16.4&2.092&74.1&18.8\\
60$^o$&356.6&2.431&70.3&21.7&2.692&77.4&16.7&2.567&74.1&18.8\\
70$^o$&387.0&2.634&70.1&21.9&2.920&77.3&16.8&2.787&74.1&18.8\\
90$^o$&411.8&2.833&70.7&21.4&3.113&77.3&16.8&2.963&73.8&19.0\\

\hline
\multicolumn{11}{c}{\textbf{q=40}}\\
\hline

30$^o$&263.6&1.378&69.6&44.9&1.571&79.2&31.0&1.490&75.1&36.1\\
45$^o$&372.9&1.977&70.3&43.6&2.226&79.0&31.3&2.115&74.9&36.4\\
60$^o$&456.9&2.451&70.8&42.7&2.726&78.4&32.0&2.817&74.9&36.4\\
70$^o$&495.8&2.676&71.0&42.4&2.958&78.4&32.0&2.817&74.9&36.4\\
90$^o$&527.6&2.857&71.2&42.1&3.154&78.5&31.8&2.926&74.7&36.7\\

\hline
\end{tabular}
\end{center}\end{table}

\newpage

\begin{table}[H]
\small
\centering
\caption{Approximate values $V_{rot}^{disk}$ and $q_{disk}$, with Ca I $\lambda 6439 \; \angstrom $ line for classical rotational broadening model ($P = 1$).}
\label{tabular:Disk}
\begin{tabular}{c|c|c||c|c||c|c|c|c}
\hline
\multirow{2}{*}{q}&\multirow{2}{*}{$K_v$ (km/s)}&\multirow{2}{*}{$V_{rot}^{Roche}$}&\multicolumn{3}{c||}{$\phi=0.25$}&\multicolumn{3}{c}{the phase-averaged values}\\
\cline{4-9}
&&&\footnotesize{FWHM, $\angstrom$}&$V_{rot}^{disk}$&$q_{disk}$&\footnotesize{FWHM, $\angstrom$}&$V_{rot}^{disk}$&$q_{disk}$\\
\hline
1& 123.58&94.77&3.940&118.86&0.68&3.863&116.53&0.7\\
2&189.03&91.58&3.829&115.50&1.33&3.729&112.46&1.39\\
5&298.45&90.19&3.792&114.37&3.08&3.673&110.76&3.28\\
7&345.10&90.23&3.804&114.74&4.14&3.675&110.82&4.46\\
10&400.29&90.46&3.823&115.32&5.65&3.685&111.13&6.16\\
20&518.85&91.24&3.864&116.56&10.44&3.717&112.10&11.54\\
40&664.93&92.18&3.903&117.74&19.62&3.755&113.25&21.83\\
\hline
\end{tabular}
\end{table}

\end{document}